\documentclass[english,reprint, superscriptaddress, breaklinks=true, showkeys, showpacs=false, nofootinbib]{revtex4}
\usepackage[T1]{fontenc}
\usepackage[latin9]{inputenc}
\usepackage{color}
\usepackage{babel}
\usepackage{float}
\usepackage{amssymb}
\usepackage{graphicx}
\usepackage[unicode=true,pdfusetitle,
 bookmarks=true,bookmarksnumbered=false,bookmarksopen=false,
 breaklinks=true,pdfborder={0 0 0},backref=false,colorlinks=true]
 {hyperref}
\usepackage{breakurl}

\makeatletter
\@ifundefined{textcolor}{}
{%
 \definecolor{BLACK}{gray}{0}
 \definecolor{WHITE}{gray}{1}
 \definecolor{RED}{rgb}{1,0,0}
 \definecolor{GREEN}{rgb}{0,1,0}
 \definecolor{BLUE}{rgb}{0,0,1}
 \definecolor{CYAN}{cmyk}{1,0,0,0}
 \definecolor{MAGENTA}{cmyk}{0,1,0,0}
 \definecolor{YELLOW}{cmyk}{0,0,1,0}
}

\hypersetup{colorlinks=true,citecolor=blue,linkcolor=cyan,urlcolor=blue,filecolor= green, breaklinks=true}
\usepackage{url}
\usepackage{breakurl}

\makeatother

\begin{document}

\title{Entanglement production by the magnetic dipolar interaction dynamics}

\author{Douglas F. Pinto}

\address{Departamento de F\'isica, Centro de Ci\^encias Naturais e Exatas, Universidade Federal de Santa Maria, Avenida Roraima 1000, 97105-900, Santa Maria, RS, Brazil}

\author{Jonas Maziero}

\email{jonas.maziero@ufsm.br}

\address{Departamento de F\'isica, Centro de Ci\^encias Naturais e Exatas, Universidade Federal de Santa Maria, Avenida Roraima 1000, 97105-900, Santa Maria, RS, Brazil}
\begin{abstract}
We consider two qubits prepared in a product state and evolved under
the magnetic dipolar interaction (MDI). We describe the dependence
of the entanglement generated by the MDI with time, with the interaction
parameters, and with the system's initial state, identifying the symmetry
and coherence aspects of those initial configurations that yield the
maximal entanglement. We also show how one can obtain maximum entanglement
from the MDI applied to some families of partially entangled initial
states.
\end{abstract}


\maketitle

\section{Introduction}

\label{intro}

In view of its possible application as a quantum channel for quantum
communication and as a resource for quantum computation tasks, the
quantum correlations in the Gibbs thermal state \cite{popescuT} associated
with the magnetic dipolar interaction (MDI) \cite{neumann,dolde,choi}
have been receiving considerable attention in the quantum information
science literature \cite{Furman2012,Kuznetsova2013,Furman2014,Castro2016}.
The dynamic behavior of entanglement and of others quantum correlations
has been investigated too \cite{Furman2008,Hu2015,Khan2016,Mohamed2013,Namitha2018}.
Besides, the MDI was used to simulate spin systems \cite{Zhou2015}
and to obtain an Ising interaction \cite{Yun2015} from which CNOT
gates (an essential ingredient for universal quantum computation \cite{Nielsen2000})
can be implemented. Due to the creation of quantum correlations between
system and environment \cite{Maziero2010c,Pozzobom2017a}, which leads
to the classicality of the first, the MDI is the source of noise is
several physical systems \cite{Klauder1962,Soares-Pinto2011a,Shiddiq2016,Ota2007,Stamp2012,Witzel2012,Annabestani2017}.
So it is important, from the fundamental and practical points of view,
to investigate the dynamics of quantum coherence and of quantum correlations
due to the MDI.

Incoherent operations, i.e., quantum operations that cannot generate
superpositions of orthogonal states from their mixtures, are one of
the basic elements of the resource theories of coherence that have
being developed in the last few years \cite{Streltsov2017}. One crucial
aspect of this development is the interplay between coherence of subsystems
and the quantum correlations of their composites, and interesting
tradeoff relations for the transformation of coherence into entanglement
by incoherent operations have been identified \cite{Streltsov2015}.
Nevertheless, although incoherent operations are the natural ones
to consider from the resource theory perspective, for practical purposes
it is also relevant to investigate the capabilities of some common
physical operations to convert initial coherence into entanglement.
In this article we shall perform that kind of investigation by regarding
the MDI.

We organized the remainder of this article in the following manner.
After presenting the regarded MDI Hamiltonian in Sec. \ref{sec:MDI},
we consider the evolved states generated by this interaction for initial
product pure (Sec. \ref{sec:pure}) or mixed (Sec. \ref{sec:mixed})
states and investigate the dependence of the transformation of local
quantum coherence into quantum entanglement by the MDI on the interaction
parameters, on time, and on the system initial states. In Sec. \ref{sec:ent}
we show how one can obtain maximum entanglement from partially entangled
states using the MDI. Our conclusions are presented in Sec. \ref{sec:conc}.

\section{Hamiltonian for the Magnetic Dipolar Interaction}

\label{sec:MDI}

The Hamiltonian for the magnetic dipolar interaction (MDI) reads (see
\cite{oliveira} and references therein): 
\begin{equation}
H=D[(\vec{\sigma}\otimes\sigma_{0})\cdot(\sigma_{0}\otimes\vec{\sigma})-3\hat{n}\cdot\vec{\sigma}\otimes\hat{n}\cdot\vec{\sigma}],
\end{equation}
with $r$ being the distance between the dipoles centers and $\hat{n}$
is a unit vector in $\mathbb{R}^{3}$ pointing from one dipole to
the other, $\sigma_{0}$ is the $2\mathrm{x}2$ identity matrix, and
$\vec{\sigma}=(\sigma_{1},\sigma_{2},\sigma_{3})$ is the vector of
Pauli matrices. The strength of the MDI is given by the distance-related
parameter $D=\mu_{0}\hbar^{2}\gamma_{a}\gamma_{b}/16\pi r^{3}$, with
$\mu_{0}$ being the vacuum permeability and $\gamma_{s}$ is the
particle $s=a,b$ gyromagnetic ratio. Throughout this paper we use
Planck's constant $\hbar=1$ and set $D=1$, which for this Hamiltonian
is equivalent to measure time in units of $D/\hbar$.

When we deal with two-level systems, it follows that $V\hat{n}\cdot\vec{\sigma}V^{\dagger}=(O\hat{n})\cdot\vec{\sigma}=\hat{n}'\cdot\vec{\sigma}$,
where $V\in SU(2)$ and $O\in SO(3)$ (see e.g. \cite{Horodecki_NF1,Horodecki_NF2}).
So, as $\sum_{j}V\sigma_{j}V^{\dagger}\otimes V\sigma_{j}V^{\dagger}=\sum_{j}\sigma_{j}\otimes\sigma_{j}$
we shall have
\begin{equation}
V\otimes VHV^{\dagger}\otimes V^{\dagger}=(\vec{\sigma}\otimes\sigma_{0})\cdot(\sigma_{0}\otimes\vec{\sigma})-3\hat{n}'\cdot\vec{\sigma}\otimes\hat{n}'\cdot\vec{\sigma}.
\end{equation}
We see thus that by changing the relative spacial orientation of the
dipoles centers ($\hat{n}\rightarrow\hat{n}'$) we will not affect
the entanglement of the MDI Hamiltonian eigenstates nor of its associated
Gibbs thermal state\footnote{The Gibbs thermal state has the form: $\rho_{th}=Z^{-1}e^{-\beta H}$,
where $Z=\mathrm{Tr}(e^{-\beta H})$ is the partition function and
$\beta=(k_{B}T)^{-1}$, with $T$ being the bath temperature and $k_{B}$
is the Boltzmann constant.}, because
\begin{equation}
e^{cV\otimes VHV^{\dagger}\otimes V^{\dagger}}=V\otimes Ve^{cH}V^{\dagger}\otimes V^{\dagger}
\end{equation}
for $c\in\mathbb{C}$. But, as we will show in this article, the dynamical
generation of entanglement by the MDI is affected by the change in
spacial orientation $\hat{n}\rightarrow\hat{n}'$, which corresponds
to a general local rotation of the dipoles initial states before their
original MDI is turned on. For simplicity, all results we shall present
hereafter are for $\hat{n}=(0,0,1)$, so that the dipoles centers
lie in the $z$ axis. In this case
\begin{eqnarray}
H & = & 2^{-1}(\sigma_{1}\otimes\sigma_{1}+\sigma_{2}\otimes\sigma_{2}-2\sigma_{3}\otimes\sigma_{3})\\
 & = & 0|\Psi_{-}\rangle\langle\Psi_{-}|+2|\Psi_{+}\rangle\langle\Psi_{+}|-|\Phi_{-}\rangle\langle\Phi_{-}|-|\Phi_{+}\rangle\langle\Phi_{+}|,
\end{eqnarray}
with $|\Psi_{\pm}\rangle=2^{-1/2}(|01\rangle\pm|10\rangle)$ and $|\Phi_{\pm}\rangle=2^{-1/2}(|00\rangle\pm|11\rangle)$
being the Bell's states. Throughout this article we use the notation
$|j\rangle\otimes|k\rangle=|jk\rangle$, where $\{|j\rangle\}_{j=0}^{1}$
is the standard basis for $\mathbb{C}^{2}$. In the next sections,
the dynamics generated by this Hamiltonian is studied with particular
focus on its capabilities to transform local quantum coherence into
quantum entanglement or partial entanglement into maximal entanglement.

\section{entanglement production by the magnetic dipolar interaction for product
initial states}

\subsection{Initial product-pure states}

\label{sec:pure}

In this subsection we consider the two dipoles prepared in a product-pure
state $|\psi_{a}\rangle\otimes|\psi_{b}\rangle$, with
\begin{equation}
|\psi_{s}\rangle=\alpha_{s}|0\rangle+\beta_{s}|1\rangle=\cos\frac{\theta_{s}}{2}|0\rangle+\sin\frac{\theta_{s}}{2}|1\rangle
\end{equation}
and $\theta_{s}\in[0,2\pi]$ for $s=a,b$, i.e., we consider two coaxial
rings in the Bloch's sphere picture for the initial states. With reference
to the standard basis $\{|0\rangle,|1\rangle\}$, the $l_{1}$-norm
quantum coherence \cite{Baumgratz2014} of such a state is:
\begin{equation}
C_{l_{1}}(|\psi_{s}\rangle)=2|\alpha_{s}|\sqrt{1-|\alpha_{s}|^{2}}=|\sin\theta_{s}|.\label{eq:icoh}
\end{equation}
For the aforementioned initial state, the evolved state under the
magnetic dipolar interaction (MDI) is given (up to a global phase)
by:
\begin{eqnarray}
|\Psi_{t}\rangle & = & U_{t}|\psi_{a}\rangle\otimes|\psi_{b}\rangle=e^{-iHt}|\psi_{a}\rangle\otimes|\psi_{b}\rangle\nonumber \\
 & = & (\alpha_{a}\beta_{b}\cos t-i\beta_{a}\alpha_{b}\sin t)|01\rangle+(\beta_{a}\alpha_{b}\cos t-i\alpha_{a}\beta_{b}\sin t)|10\rangle+e^{i2t}(\alpha_{a}\alpha_{b}|00\rangle+\beta_{a}\beta_{b}|11\rangle).\label{eq:psit}
\end{eqnarray}

\begin{figure}[H]
\begin{raggedright}
\includegraphics[scale=0.85]{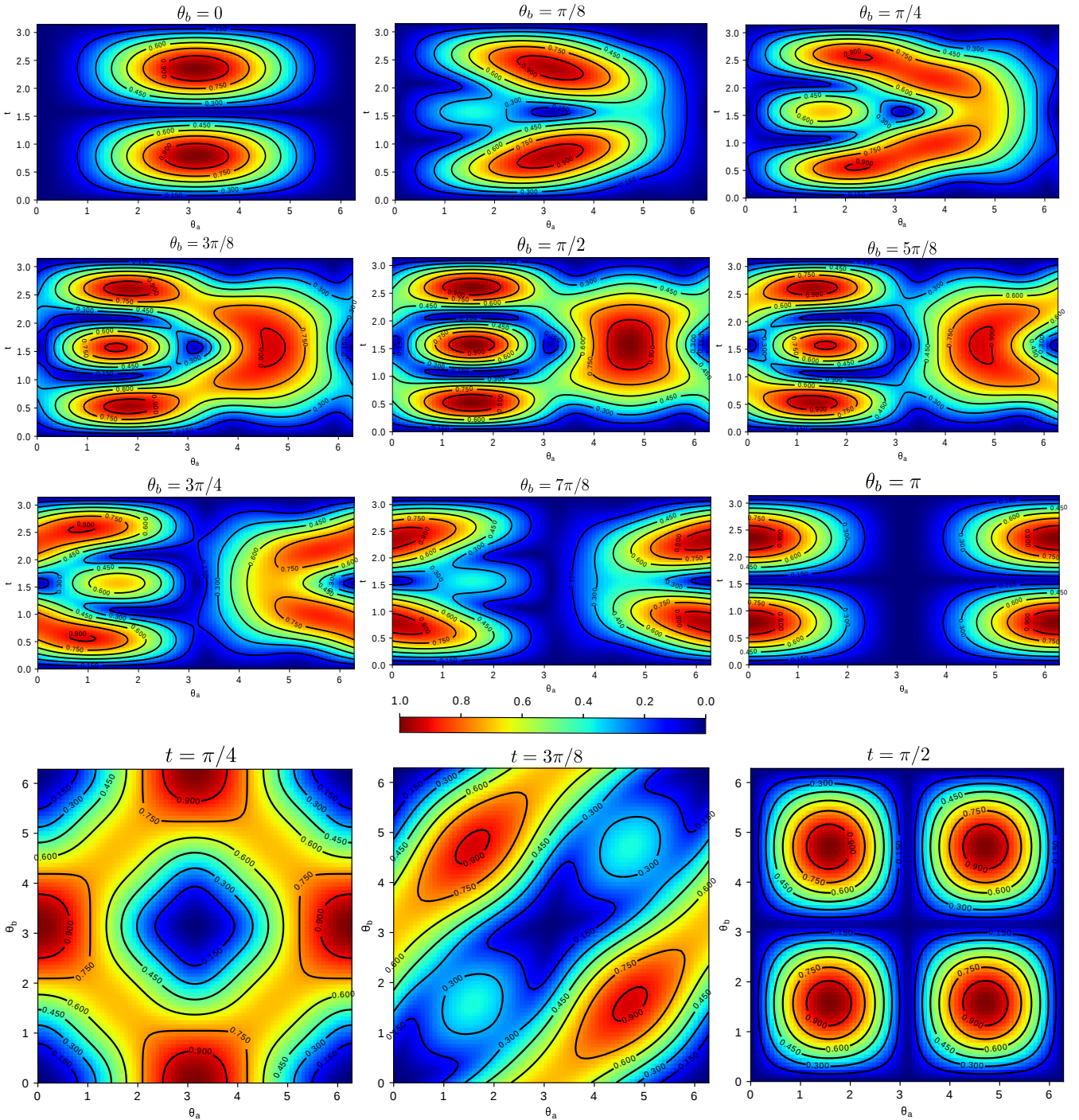}
\par\end{raggedright}

\caption{(color online) In the \emph{first three rows} of plots is shown the
entanglement as a function of time (in units of $D/\hbar$) and of
dipole $a$ initial state for some initial states of dipole $b$.
We verified that entanglement is a periodic function in time with
period $\pi$. Besides, the plots for $\theta_{b}=\pi+\phi$, for
$\phi\in[0,\pi]$, look just like those for $\theta_{b}=\pi-\phi$
reflected in relation to the $\theta_{a}=\pi$ axis. In the \emph{last
row} of plots is shown the entanglement generated by the MDI as a
function of the angles that determine the dipoles initial states for
some values of time. The figure for $t=\pi/8$ is equal to that for
$t=3\pi/8$ rotated clockwise in the $\theta_{a}\mathrm{x}\theta_{b}$
plane by $\pi/2$. For the other values of time, the values of entanglement
are equal or lesser than the corresponding values in these four figures.
Overall, the dependence of the entanglement generated by the MDI on
the initial local coherences is far from simple, and we highlight
its most important characteristics in the main text. }

\label{fig:Extaxt}
\end{figure}

In this subsection we shall compute the entanglement of the evolved
state in Eq. (\ref{eq:psit}) using the concurrence \cite{Wootters1998},
which for the pure state above is:
\begin{eqnarray}
 &  & E_{C}(|\Psi_{t}\rangle)=|\langle\Psi_{t}|\sigma_{2}\otimes\sigma_{2}|\Psi_{t}^{*}\rangle|=\sqrt{f^{2}+g^{2}},
\end{eqnarray}
with $f=2\alpha_{a}\beta_{a}\alpha_{b}\beta_{b}(\cos2t-\cos4t)$ and
$g=(\alpha_{a}^{2}\beta_{b}^{2}+\beta_{a}^{2}\alpha_{b}^{2})\sin2t+2\alpha_{a}\beta_{a}\alpha_{b}\beta_{b}\sin4t,$
where $|\Psi_{t}^{*}\rangle$ is the complex conjugate of $|\Psi_{t}\rangle$
represented in the standard basis. Some examples of the time and initial
state dependence of the entanglement created by the MDI are shown
graphically in Fig. \ref{fig:Extaxt}. If it would to be possible
to experimentally turn off the MDI at any given instant of time, we
could choose the moment at which the two dipoles share the greater
value of entanglement. So, in Fig. \ref{fig:Extaxt} we present also
the dependence of the entanglement generated by the MDI on the angles
$\theta_{a}$ and $\theta_{b}$ for some fixed values of time.

We see in Fig. \ref{fig:Extaxt} that only the following set of initial
states $\{|01\rangle,|10\rangle,|++\rangle,|+-\rangle,|-+\rangle,|--\rangle\},$
with $|\pm\rangle=2^{-1/2}(|0\rangle\pm|1\rangle)$, yields the maximum
possible value of entanglement. While the last four initial states
are maximally coherent for the two dipoles, the first two initial
configurations have zero local coherence. We notice then that the
MDI is not an incoherent operation, since it can produce entanglement
from incoherent states (see e.g. Ref. \cite{Streltsov2015}). On the
other hand, we can understand the non-equivalence between the two
pairs of incoherent states ($|01\rangle$, $|10\rangle$) and ($|00\rangle$,
$|11\rangle$) with regard to entanglement generation by noticing
that as $[H,\sigma_{3}\otimes\sigma_{0}+\sigma_{0}\otimes\sigma_{3}]=0$
the dynamics under the MDI conserves the total number of excitations
of the system. So, the later pair of states remain confined to their
subspaces, which involves only the product states, while the first
pair can superpose to produce entanglement.

The commutation relation above can be used also to show that $U_{t}$
commutes with $R_{z}(\delta)\otimes R_{z}(\delta)$, where $R_{z}(\delta)=\exp(-i\delta\sigma_{3}/2)$.
Once $E_{C}(R_{z}(\delta)\otimes R_{z}(\delta)U_{t}|\psi_{a}\rangle\otimes|\psi_{b}\rangle)=E_{C}(U_{t}R_{z}(\delta)|\psi_{a}\rangle\otimes R_{z}(\delta)|\psi_{b}\rangle),$
our main conclusions about the coherence-entanglement conversion by
the MDI shall be the same for any orientation we use for the two coaxial
rings of initial states. We emphasize e.g. that any pair of ``parallel''
or ``anti-parallel'' states in the equator of the Bloch sphere shall
lead to maximal entanglement if evolved under the MDI. 

Now that we have presented these general results for the entanglement
generated by the MDI for the initial spacial orientation $\hat{n}=(0,0,1)$,
we can show explicitly that although the MDI eigenstates and thermal
entanglement do not change by changing the dipole centers spatial
orientation, the dynamical generation of non-separable states by the
MDI can be greatly affected by this kind of operation. The main point
here is that the change $\hat{n}\rightarrow\hat{n}'$ is equivalent
to modifying the evolution operator as $e^{-iHt}\rightarrow V\otimes Ve^{-iHt}V^{\dagger}\otimes V^{\dagger}$,
which is effectively equivalent, with respect to entanglement generation,
to change the initial state to $V^{\dagger}|\psi_{a}\rangle\otimes V^{\dagger}|\psi_{b}\rangle$.
As an extreme example, let us consider the change $\hat{n}\rightarrow\hat{n}'$
corresponding to the unitary operation $V^{\dagger}$ that leads to
a rotation of the initial states Bloch vectors by $\pi/2$ around
the $y$ axis. This rotation applied to the initial state $|00\rangle$
returns the state $|++\rangle$, and in this case we would go from
a situation where no entanglement is created to another initial state
that gives us maximal entanglement by the MDI. Of course, this issue
will appear also for the initial mixed-product states we study in
the next subsection.

The results presented in this section indicate the non-existence of
a direct-general temporal correlation between the values of coherence
and entanglement. But, for completeness, we present in the Appendix
the time evolution of local quantum coherence in this case.

\subsection{Initial product-mixed states}

\label{sec:mixed}

In order to investigate the effect of the purity of the initial state
on the entanglement produced by MDI, we regard as initial states the
following product states of the two dipoles: $\rho_{ja}\otimes\rho_{jb}$,
where $\rho_{js}=2^{-1}(\sigma_{0}+r_{js}\sigma_{j})$ with $r_{js}=\mathrm{Tr}(\rho_{s}\sigma_{j})\in[-1,1]$
and $j=1$ or $j=3$ (these are, respectively, the $x$ and $z$ axis
in the Bloch's ball). For these local states, the $l_{1}$-norm coherence
is given by $C_{l_{1}}(\rho_{1s})=|r_{1s}|$ and $C_{l_{1}}(\rho_{3s})=0$,
i.e., we use a generally coherent or an incoherent initial state.
The local purities read $P(\rho_{js})=\mathrm{Tr}(\rho_{js}^{2})=2^{-1}(1+r_{js}^{2})$.
Notice that for both states $\rho_{js}$ the purity is a monotonously
increasing function of $|r_{js}|$.

Here the evolved states, $\rho_{j}=e^{-iHt}(\rho_{ja}\otimes\rho_{jb})e^{iHt}$,
read
\begin{eqnarray}
4\rho_{3} & = & (1+r_{3a})(1+r_{3b})|00\rangle\langle00|+[1-r_{3a}r_{3b}+(r_{3a}-r_{3b})\cos2t]|01\rangle\langle01|+i(r_{3a}-r_{3b})\sin2t|01\rangle\langle10|\label{eq:rho3}\\
 &  & -i(r_{3a}-r_{3b})\sin2t|10\rangle\langle01|+[1-r_{3a}r_{3b}-(r_{3a}-r_{3b})\cos2t]|10\rangle\langle10|+(1-r_{3a})(1-r_{3b})|11\rangle\langle11|\nonumber 
\end{eqnarray}
and
\begin{eqnarray}
4\rho_{1} & = & (1-r_{1a}r_{1b})(|\Psi_{-}\rangle\langle\Psi_{-}|+|\Phi_{-}\rangle\langle\Phi_{-}|)+(1+r_{1a}r_{1b})(|\Psi_{+}\rangle\langle\Psi_{+}|+|\Phi_{+}\rangle\langle\Phi_{+}|)\label{eq:rho1}\\
 &  & +(r_{1b}-r_{1a})(e^{it}|\Phi_{-}\rangle\langle\Psi_{-}|+e^{-it}|\Psi_{-}\rangle\langle\Phi_{-}|)+(r_{1b}+r_{1a})(e^{i3t}|\Phi_{+}\rangle\langle\Psi_{+}|+e^{-i3t}|\Psi_{+}\rangle\langle\Phi_{+}|).\nonumber 
\end{eqnarray}

\begin{figure}[H]
\includegraphics[scale=0.64]{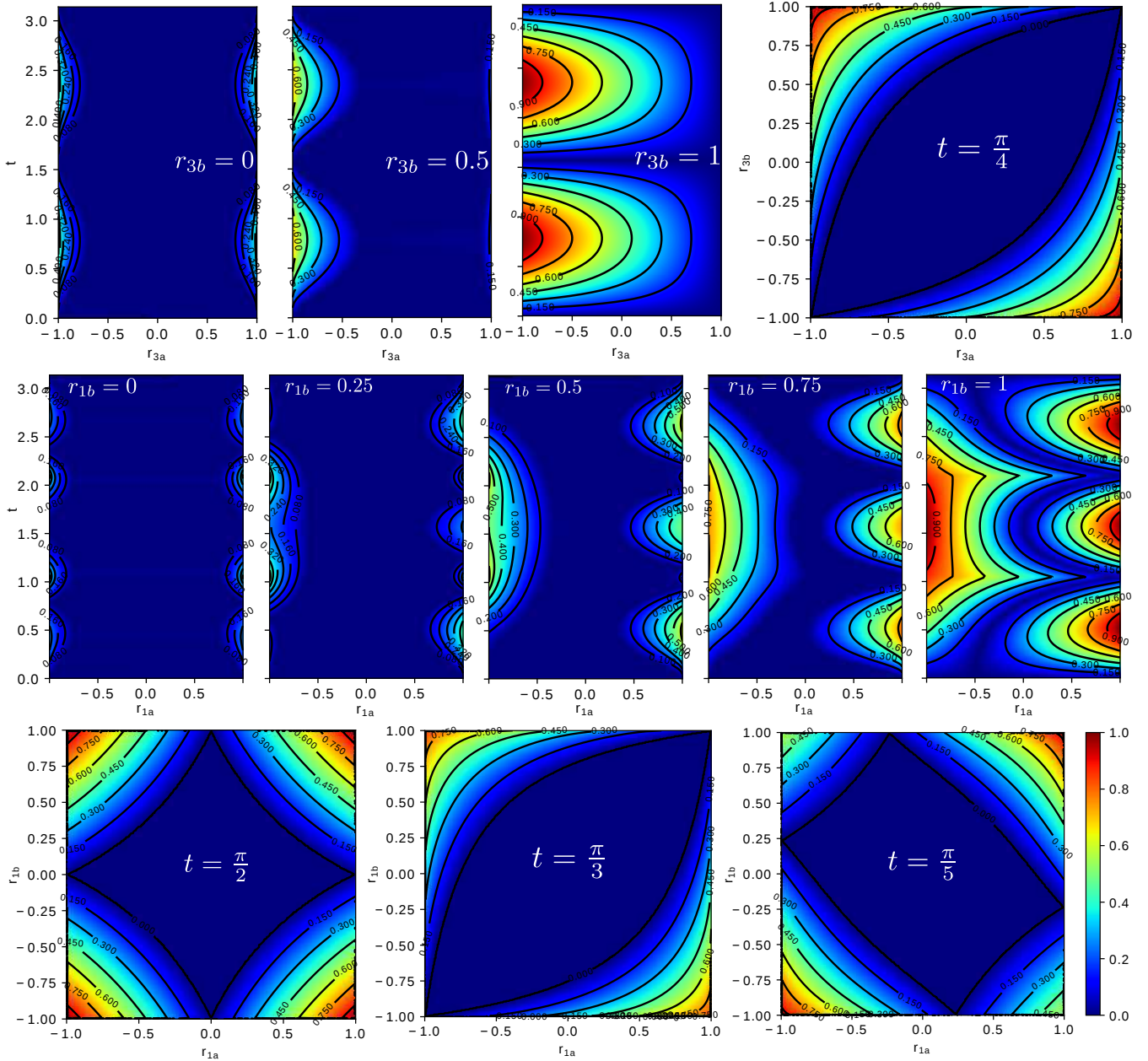}

\caption{(color online) In the \emph{first row} of plots is presented the entanglement
of $\rho_{3}$. The first three figures show the simple temporal dependence
of $E_{C}$ with time in this case. The figures for $r_{3b}=-x$ look
just like those for $r_{3b}=x$ reflected in relation to the $r_{3a}=0$
axis. As in the last subsection, the period of $E_{C}$ in $t$ is
equal to $\pi$. In the last plot in this row is shown the entanglement
generated by the MDI for $t=\pi/4$. Actually, maximum entanglement
is obtained in this case for all $t=(2n+1)\pi/4$ with $n\in\mathbb{N}$.
In the \emph{second and third rows} of plots is shown the entanglement
of the state $\rho_{1}$. The temporal dependence of $E_{C}$ in this
case (figures in the second row) is more involved than for $\rho_{3}$.
However, here also we have the pattern for $E_{C}$ for $r_{1b}=-x$
equivalent to that for $r_{1b}=x$ reflected in relation to the $r_{1a}=0$
axis. Although for $\rho_{1}$ we cannot identify instants in time
giving the maximum entanglement in general, in the last three figures
we show $E_{C}$ as a function of the initial states for three times
that should contribute the most for that general maximum.}

\label{fig:ERho}
\end{figure}

For bipartite mixed states of two qubits the entanglement concurrence
is computed using \cite{Wootters1998}:
\begin{equation}
E_{C}(\rho)=\max(0,\sqrt{\lambda_{1}}-\sqrt{\lambda_{2}}-\sqrt{\lambda_{3}}-\sqrt{\lambda_{4}}),
\end{equation}
with $\{\lambda_{j}\}_{j=1}^{4}$ being the eigenvalues of $\rho\sigma_{2}\otimes\sigma_{2}\rho^{*}\sigma_{2}\otimes\sigma_{2}$
indexed in decreasing order and $\rho^{*}$ is the complex conjugate
of the system's density matrix $\rho$. In Fig. \ref{fig:ERho} we
show the numerical results for{\tiny{} }the entanglement concurrence
of $\rho_{3}$ and of $\rho_{1}$ as a function of time and of dipoles
$a$ and $b$ initial states. 

As one can observe in Fig. \ref{fig:ERho}, the entanglement has an
oscillatory behavior with time and $E_{C}$ generally increases with
the total purity of the dipoles initial states. This proportionality
is confirmed by the maximum values of the entanglement as a function
of the dipoles initial states. We observe that for $r_{ja}=r_{jb}=0$
the initial state is proportional to the identity and no entanglement
is created. Besides, there is a minimal total purity of the dipoles
below which we get no entanglement. Of course, in the limiting cases
of maximum purity, coinciding with those of the last subsection, the
MDI produce the maximum possible amount of entanglement. Notwithstanding,
as we have shown here, although purity is a important figure to consider
regarding the entanglement of the evolved state, the symmetry of the
initial state with relation to the Hamiltonian generating the evolution
is also relevant for analyzing the dynamical creation of entanglement.

Once more, because $U_{t}$ commutes with $R_{z}(\delta)\otimes R_{z}(\delta)$,
the results presented in this subsection shall be valid for all initial
states of the two qubits corresponding to parallel axes in the $xy$
plane of the Bloch sphere.

\section{entanglement production by the magnetic dipolar interaction for partially
entangled initial states}

\label{sec:ent}

\begin{figure}[H]
\begin{centering}
\includegraphics[scale=0.83]{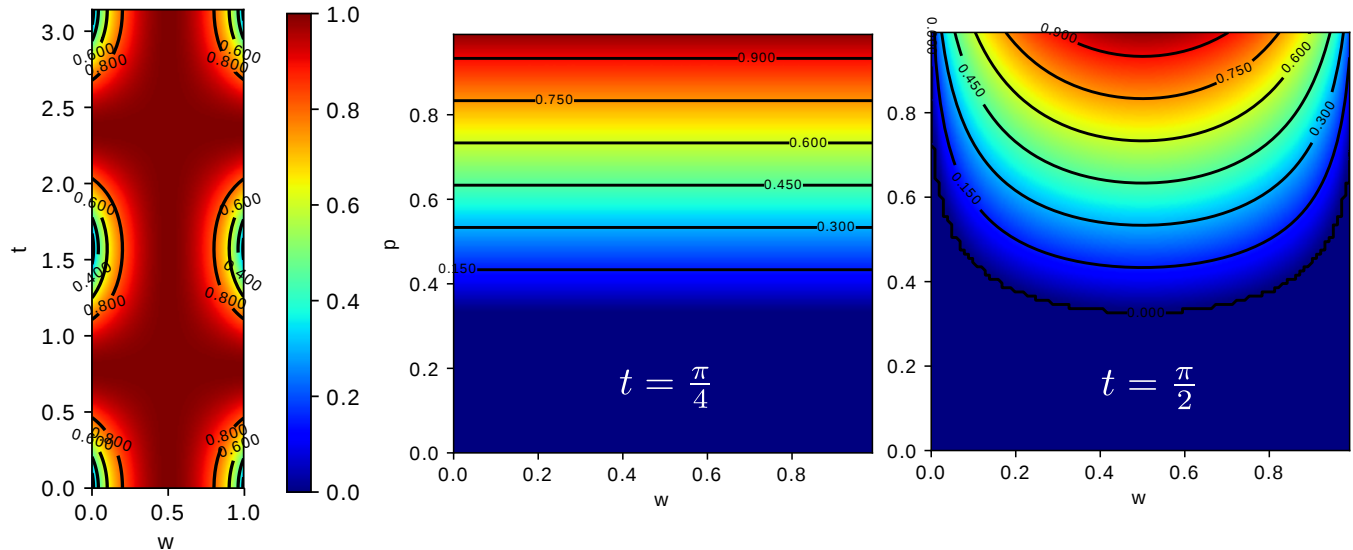}
\par\end{centering}

\caption{(color online) On the \emph{first} plot is shown the entanglement
generated by the MDI as a function of time (in units of $D/\hbar$)
and of the parameter $w$, which determines the partially entangled
initial state. The two plots on the \emph{right} show the entanglement
dependence on the depolarization parameter $p$ for two instants of
time. }

\label{fig:Ew}
\end{figure}

In this section we shall study partially entangled states evolving
under the magnetic dipolar interaction (MDI). As the computational
base states $|00\rangle$ and $|11\rangle$ gain the same phase when
evolved under the MDI, we shall start by regarding the initial pure
state
\begin{equation}
|\Psi_{0}\rangle=\sqrt{w}|01\rangle+\sqrt{1-w}|10\rangle
\end{equation}
with $w\in[0,1]$. Actually, we can get $|\Psi_{0}\rangle$ from superpositions
of $|00\rangle$ and $|11\rangle$ by applying the flip operation
$\sigma_{1}$ to one of the dipoles before they interact. For the
initial state $|\Psi_{0}\rangle$, the evolved state reads, up to
a global phase, as follows
\begin{equation}
|\Psi_{t}\rangle=(\sqrt{w}\cos t-i\sqrt{1-w}\sin t)|01\rangle+(\sqrt{1-w}\cos t-i\sqrt{w}\sin t)|10\rangle.\label{eq:psit_E}
\end{equation}
The entanglement concurrence of this pure state is given by
\begin{equation}
E_{C}(|\Psi_{t}\rangle)=\sqrt{\sin^{2}2t+4w(1-w)\cos^{2}2t},
\end{equation}
and is shown graphically in Fig. \ref{fig:Ew}. We notice in this
figure that for any value of the entanglement of the initial state,
there will be points in time for which the maximum value for the entanglement
is attained. Actually, we see that the equation $E_{C}(|\Psi_{t}\rangle)=1$
is satisfied for any value of $w$ if $t=(2n+1)\pi/4$ with $n\in\mathbb{N}$.
So, if the MDI between the qubits is turned off in any of these instants
of time, we shall have prepared a maximally entangled state from any
of the partially entangled or product states investigated in this
section. 

To give an example of the effect of decreasing the purity of the initial
state, let us consider $|\Psi_{0}\rangle$ subject to the depolarization
channel \cite{Nielsen2000}, whose action is leaving a state alone
with probability $p$ or turning it into the maximal uncertain state
with probability $1-p$, i.e., $|\Psi_{0}\rangle\rightarrow\rho_{d}=(1-p)2^{-2}\sigma_{0}\otimes\sigma_{0}+p|\Psi_{0}\rangle\langle\Psi_{0}|$.
For this initial state, the evolved state under the MDI reads:
\begin{equation}
\rho_{t}=U_{t}\rho_{d}U_{t}^{\dagger}=(1-p)2^{-2}\sigma_{0}\otimes\sigma_{0}+p|\Psi_{t}\rangle\langle\Psi_{t}|.
\end{equation}
The entanglement of this state is shown in Fig. \ref{fig:Ew} for
some instants of time. As expected, the entanglement of $\rho_{t}$
is that of $|\Psi_{t}\rangle$ diminished proportionally to $1-p$;
and there are values of $p$ below which no entanglement is generated
by the MDI. 

\section{Conclusions}

\label{sec:conc}

Entanglement is an important resource in quantum information science,
being essential for quantum teleportation \cite{bennett,popescu,Castro2016,cavalcanti}
and for its applications in quantum networks \cite{pirandola} and
quantum computation \cite{gottesman}. In this article we investigated
the capabilities of the magnetic dipolar interaction to generate entanglement.
The MDI is a coherent operation that was shown to be capable of generating
maximally entangled states from local maximally coherent or incoherent
states. The symmetry of the initial state with relation to the MDI
Hamiltonian was identified as a determinant property regarding entanglement
production, besides the initial states coherences and/or purities.
Finally, we identified conditions under which some classes of partially
entangled initial states can be transformed into maximally entangled
states by the MDI. We believe that the interesting dynamical properties
of the MDI reported in this article can contribute to its deployment
in quantum information science.
\begin{acknowledgments}
This work was supported by the Brazilian National Institute for the Science and Technology of Quantum Information (INCT-IQ), process 465469/2014-0.
\end{acknowledgments}

\appendix*

\section{Dynamics of local quantum coherence for initial pure-product states}

\label{app1}

\begin{figure}
\begin{centering}
\includegraphics[scale=0.81]{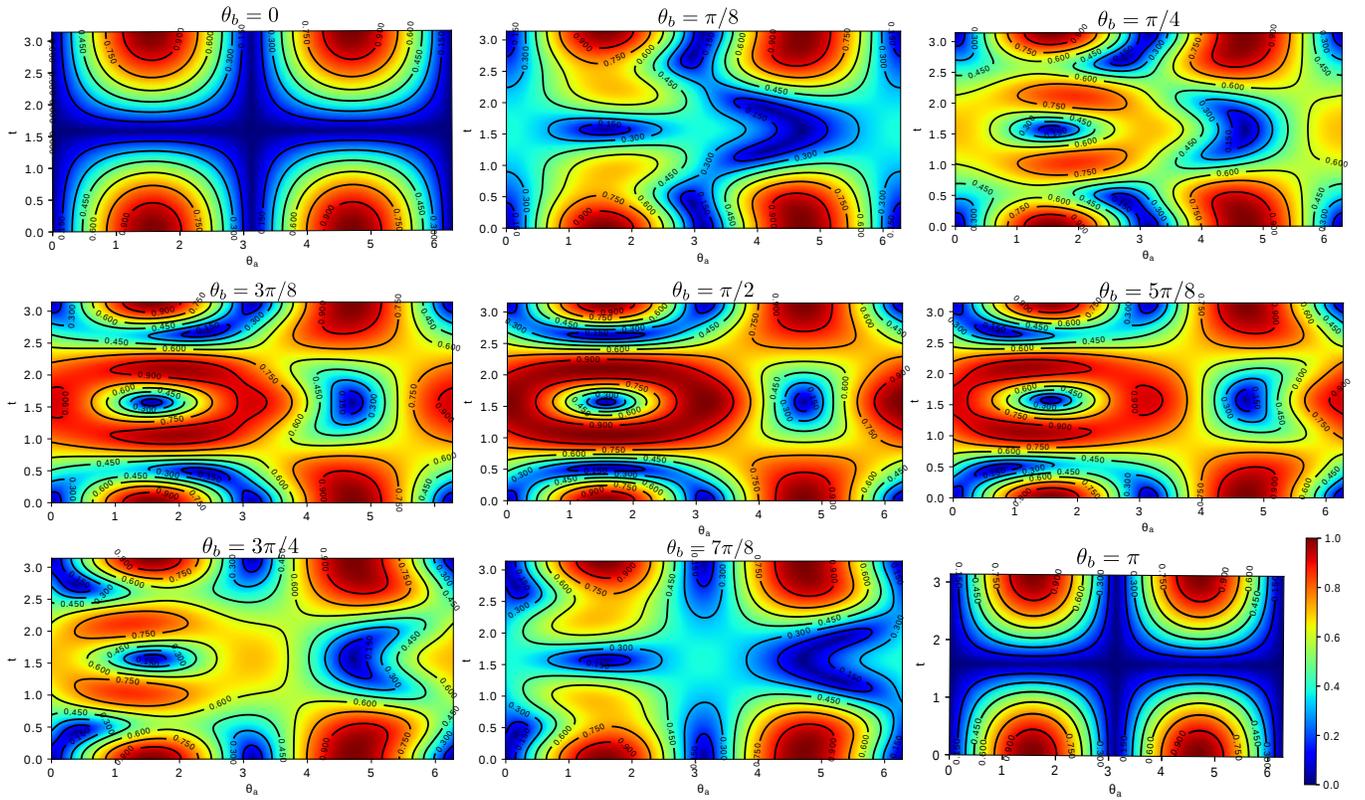}
\par\end{centering}

\caption{(color online) Local quantum coherence for initial pure-product states
as function of time (in units of $D/\hbar$) and of dipole $a$ initial
state for some initial states of dipole $b$.}

\label{fig:CtPsi}
\end{figure}

Here we use the $l_{1}$-norm coherence \cite{Baumgratz2014}, $C(\rho)=\sum_{j\ne k}|\langle j|\rho|k\rangle|$,
to quantify quantum coherence. By taking the partial trace \cite{MazieroPTr}
over one of the two dipoles, whose composite state is (\ref{eq:psit}),
we obtain the reduced density operator $\rho_{r}=\mathrm{Tr}_{p}(|\Psi_{t}\rangle\langle\Psi_{t}|)$.
The quantum coherence of this state reads
\begin{eqnarray}
2^{-2}C^{2}(\rho_{r}) & = & \alpha_{a}^{2}\beta_{a}^{2}(\alpha_{b}^{4}+\beta_{b}^{4})\cos^{2}t+\alpha_{b}^{2}\beta_{b}^{2}(\alpha_{a}^{4}+\beta_{a}^{4})\sin^{2}t+2\alpha_{a}^{2}\beta_{a}^{2}\alpha_{b}^{2}\beta_{b}^{2}\cos2t\cos4t\nonumber \\
 &  & -\alpha_{a}\beta_{a}\alpha_{b}\beta_{b}(\alpha_{a}^{2}\beta_{b}^{2}+\beta_{a}^{2}\alpha_{b}^{2})\sin2t\sin4t.
\end{eqnarray}
In Fig. \ref{fig:CtPsi} we plot this quantity as a function of time
and of the dipole $a$ initial state for some initial states of dipole
$b$. Comparison with Fig. \ref{fig:Extaxt} confirms the non-existence
of a general temporal correlation between the values of coherence
and entanglement. 

\bibliographystyle{apsrev4-1}
\bibliography{dipolar_cohent}

\end{document}